\documentclass[sigconf,10pt]{acmart}
\AtBeginDocument{%
  \providecommand\BibTeX{{%
    \normalfont B\kern-0.5em{\scshape i\kern-0.25em b}\kern-0.8em\TeX}}}

\setcopyright{acmcopyright}
\copyrightyear{2018}
\acmYear{2018}
\acmDOI{XXXXXXX.XXXXXXX}

\acmConference[Conference acronym 'XX]{Make sure to enter the correct
  conference title from your rights confirmation emai}{June 03--05,
  2018}{Woodstock, NY}
\acmPrice{15.00}
\acmISBN{978-1-4503-XXXX-X/18/06}

\usepackage[ruled,linesnumbered,vlined]{algorithm2e}
\usepackage{booktabs} 
\usepackage{subcaption} 
\usepackage{listings}
\usepackage{hyperref}
\usepackage{url}
\usepackage{xcolor}
\usepackage{pdflscape}
\definecolor{mycolor}{RGB}{161,216,230}

\begin{document}
\color{black}
\title[Expanding the Scope of DAWN: A Novel Version for Weighted Shortest Path Problem]{Expanding the Scope of DAWN: A Novel Version for Weighted Shortest Path Problem}             

\author{Yelai Feng}
\orcid{0000-0002-3134-9227}             
\affiliation{
	\department{College of Electronic Engineering}              
	\institution{National University of Defense Technology}            
	\streetaddress{Huangsha Road 460}
	\city{Hefei}
	\state{Anhui}
	\postcode{230000}
	\country{CHINA}                    
}
\email{fengyelai@nudt.edu.cn}          

\renewcommand{\shortauthors}{}
\begin{abstract}
	The shortest path problem is a typical problem in graph theory with wide potential applications. The state-of-the-art single-source shortest paths algorithm on the weight graph is the  $\Delta$-stepping algorithm, which can efficiently process weighted graphs in parallel. DAWN is an algorithm that addresses the shortest path problem on unweighted graphs, and we propose a weighted version that can handle graphs with weights edges, while maintaining the high scalability and parallelism features as DAWN. The novel version requires $O(\mu m)$ and $O(\mu \cdot E_{wcc})$ times on the connected and unconnected graphs for SSSP problems, respectively. $E_{wcc}$ denote the number of edges included in the largest weakly connected component, and $\mu$ is a constant denoting the average number of path transformations in the tasks. We tested the weighted version on the real graphs from Stanford Network Analysis Platform and SuiteSparse Matrix Collection, which outperformed the solution of  $\Delta$-stepping algorithm from Gunrock, achieving a speedup of 32.285$\times$.

\end{abstract}




\maketitle

\section{Introduction}
The shortest path problem is a typical problem in graph theory, and has attracted extensive attention of researchers \cite{martin2009cuda,ortega2013new,davidson2014work,busato2015efficient,wang2019sep,wang2021fast}. The state-of-the-art method on weighted graphs is the  $\Delta$-stepping implementation and Part-Floyd algorithm \cite{meyer2003delta,djidjev2014efficient}, for SSSP (single-source shortest paths) and APSP (all-pairs shortest paths) problem, respectively. 

The  $\Delta$-stepping implementation of Dijkstra's algorithm requires $O(n+m+d\cdot L)$ times\cite{meyer2003delta}, where $d$ represents the maximum node degree and $L$ denotes the maximum shortest path weight from the source node $s$ to any node reachable from $s$. 

Feng et al. proposed a novel shortest path algorithm named DAWN (Distance Assessment algorithm With matrix operations on Networks) \cite{feng2023novel}, which require $O(m)$ and $O(E_{wcc})$ times on the connected and unconnected graphs for SSSP problems, respectively. DAWN achieves a better time complexity than the BFS algorithm, and is the state-of-the-art SSSP algorithm on unweighted graphs. However, the initial version was limited to unweighted graphs, and enabling DAWN to operate on weighted graphs has remained a pressing challenge.

We propose a version of DAWN running on weighted graphs. The weighted version of DAWN requires $O(\mu m)$ and $O(\mu \cdot E_{wcc})$ times on the connected and unconnected graphs for SSSP problems, respectively. The weighted version can also process APSP tasks and requires $O(\mu \cdot n \cdot m)$ and $O(\mu \cdot S_{wcc} \cdot E_{wcc})$ time on the connected and unconnected graphs. $S_{wcc}$ and $E_{wcc}$ denote the number of nodes and edges included in the largest WCC (Weakly Connected Component) in the graphs. $\mu$ is a constant denoting the average number of path transformations in the tasks.

The main contributions of this work are as follows:
\begin{enumerate}
	\item   We propose an efficient version of DAWN on weighted graphs, which requires $O(\mu m)$ and $O(\mu \cdot E_{wcc})$ times on the connected and unconnected graphs for SSSP problems, respectively.
	\item   We propose GSVM (General Sparse Vector-Matrix Operation) as a replacement for SOVM (Sparse Optimized Boolean Vector-Matrix Operation) implementation of DAWN. The SOVM implementation was only applicable to unweighted graphs, and GSVM improves upon the SOVM implementation, which is applied to the weighted graphs. GSVM requires $O(\frac{Step(s)}{2}m)$ times for the SSSP tasks on weighted graph, where $Step(s)$ represents the maximum steps of the shortest path from node $s$.
	\item  We propose an optimization method GOVM (General Optimized Sparse Vector-Matrix Operation) that reduces the time complexity of GSVM from $O(\frac{Step(s)}{2}m)$ to $O(\mu \cdot E_{wcc})$, while maintaining its high parallelism and scalability. If applied to an unweighted graph, each step of the GOVM implementation yields a deterministic optimal solution, resulting in $\mu = 1$ ($\mu > 1$ on the weighted graphs), meaning that each shortest path is discovered exactly once and does not change thereafter. 
	\item  DAWN is capable of solving the APSP and SSSP problems on graphs with negative weights, and can automatically exclude the influence of negative weight cycles. Prior to this, the Bellman-Ford and Floyd algorithms required $O(nm)$ and $O(n^3)$ time, respectively. In contrast, the weighted version of DAWN requires only $O(\mu \cdot E_{wcc})$ and $O(\mu \cdot S_{wcc}\cdot E_{wcc})$ time to solve the APSP and SSSP problems, respectively.
	\item   DAWN based on GOVM achieves average speedup of 32.285$\times$ and 745.834$\times$ over  $\Delta$-stepping implementation of Dijkstra's algorithm from Gunrock, for APSP and MSSP (Multi-source shortest path) tasks, respectively.

\end{enumerate}
In Section 2, we review existing approaches to solving the shortest path problem. In Section 3, we present the design of GSVM and propose the GOVM optimization method. In Section 4, we evaluate the performance of the weighted version of DAWN through a series of multidimensional comparative experiments. In Section 5, we conclude the work of this paper. We adopt the symbols which has already defined in \cite{feng2023novel}, and list the new notations as follows \ref{tab:notations}.

\begin{table}[htpb]
	\begin{center}
		\begin{minipage}{0.5\textwidth}
			\caption{Definition of notations}\label{tab:notations}%
			\begin{tabular*}{\textwidth}{@{\extracolsep{\fill}}cl@{\extracolsep{\fill}}}
				\hline
				Notation & Definition	  \\
				\hline
				$V, E$& Set of nodes and edges\\
				$v, e$& A node and edge\\
				$\mu$& Average number of path update \\
				$E_{wcc}$& maximum edge count of the largest WCC\\
				$S_{wcc}$& maximum node count of the largest WCC\\
				$E_{wcc}(i)$& Edge count of the largest WCC of node $i$\\
				$S_{wcc}(i)$& Node count of the largest WCC of node $i$\\
				$Step(s)$& Maximum step of the shortest path from node \\
				& $s$ to any node reachable from $s$\\
				$Step_{max}$& Maximum step of the shortest path in the graph\\
				$p$& Average connected probability\\
				\hline
			\end{tabular*}
		\end{minipage}
	\end{center}
\end{table}

\section{Related Works}
\label{relatedworks}
In this section, we review existing approaches to solving the shortest path problem. We will introduce the Dijkstra's algorithm, the Bellman--Ford algorithm, the Floyd algorithm, and DAWN, which are typical algorithms for solving the shortest path problems.

\subsection{Dijkstra's algorithm}

In 1959, Dijkstra et al. proposed and solved two graph problems: construct the tree of the minimum total length between $n$ nodes, and find the path of the minimum total length between two given nodes, $v_i$ and $v_j$ \cite{dijkstra1959note}. The main optimization methods of Dijkstra algorithm are priority queue optimization, binary heap optimization, and Fibonacci heap optimization \cite{johnson1977efficient,cherkassky1996shortest,raman1996priority,raman1997recent,thorup1999undirected,thorup2000ram}. 

Meyer et al. proposed an optimized Dijkstra algorithm, which can be a parallel setting for a large class of graphs and requires $O(n+m+d\cdot L)$ times \cite{meyer2003delta}. The best parallel version of the $\Delta$-stepping Dijkstra algorithm takes $O(d\cdot L\cdot \log n + \log^2 n)$ time and $O(n + m + d\cdot L \cdot \log n)$ work on average. Davidson et al. proposed an adaptation of Dijkstra's algorithm named the Near--Far implementation, which uses two buckets (Near and Far) implement an approximate priority queue and makes several simplifications to the worklist \cite{davidson2014work}. 

Wang et al. improved the Near-Far implementation by proposing a new approach called ADDS \cite{wang2021fast}, which employs multiple buckets and a sophisticated work scheduling scheme to enhance both work efficiency and parallelism. The work of ADDS demonstrates that data structures which may appear inapposite, such as priority queues, can be implemented efficiently for GPUs if an appropriate software architecture is employed \cite{wang2021fast}.

\subsection{Bellman--Ford algorithm}

Bellman et al. in 1958 proposed a novel shortest path algorithm \cite{bellman1958routing} that is suitable for sparse graphs and can tackle negative weighted graphs. The Bellman--Ford algorithm necessitates $n-1$ relaxation operations on $m$ edges, requiring $O(nm)$ time. Compared to the Dijkstra algorithm, the Bellman--Ford algorithm lends itself better to parallelism \cite{busato2015efficient,surve2017parallel}.

$\Delta$-stepping algorithm maintains eligible nodes with tentative distances in an array of buckets each of which represents a distance range of size $\Delta$ \cite{meyer2003delta}. It is a trade-off of Dijkstra and Bellman-Ford algorithm, when set with $\Delta = 1$, produces a variant of Dijkstra's algorithm, while setting $\Delta = \infty$ yields the Bellman-Ford algorithm. By adjusting $\Delta$ within the range of $[1, \infty]$, we obtain a spectrum of algorithms with varying degrees of processing time and parallelism \cite{busato2015efficient}. Wang et al. utilized the technique of dynamic delta values in developing ADDS to enhance the parallelism of the algorithm \cite{wang2021fast}.

\subsection{Floyd--Warshall algorithm}
In 1962, Floyd proposed an algorithm that uses dynamic programming to solve the shortest path problem and is easy to implement \cite{floyd1962algorithm}. The Floyd algorithm uses only two-dimensional arrays, compared to the Dijkstra algorithm, which uses sophisticated data structures for optimization.

Djidjev et al. proposed a Partitioned APSP algorithm (Part-APSP) based on the divide-and-conquer strategy, which divides the graph into various components and runs Floyd--Warshall algorithm respectively \cite{djidjev2014efficient}. And Part-APSP requires $O(n^{2.25})$ time on the planar graphs. Yang et al. propose a novel solution named Fast APSP algorithm \cite{10.1145/3524059.3532365}, 
which store the whole graph as a compressed storage format in each process of the distributed computing clusters. In contrast to the Part-APSP, Fast APSP algorithm combine the Floyd algorithm with the Dijkstra algorithm,which uses local Floyd and global Dijkstra algorithms simultaneously \cite{10.1145/3524059.3532365}.

Piyush Sao et al. proposed a optimized Floyd--Warshall algorithm that computes the APSP for GPU accelerated clusters \cite{10.1145/3431379.3460651}. To achieve high parallel efficiency, they address two challenges: reducing high communication overhead and addressing limited GPU memory.

\subsection{DAWN}
Feng et al. propose a the high scalability and parallelism shortest path algorithm named DAWN (Distance Assessment algorithm With matrix operations on Networks) \cite{feng2023novel}, which is the state-of-the-art SSSP algorithm on unweighted graphs. Feng et al. proposed two implementations for DAWN, namely BOVM and SOVM. On dense graphs, DAWN based on BOVM can complete the APSP task in near $O(n^2)$ time. On general graphs, DAWN based on SOVM requires $O(m)$ and $O(E_{wcc})$ time on the connected and unconnected graphs for SSSP problems, respectively.

\section{Methods}
In this section, we introduce the principle of weighted version of DAWN, then shown the design of general sparse vector-matrix operation and propose optimization of GSVM. 

Furthermore, we will discuss about the differences of DAWN based on GOVM, Dijkstra's algorithm and Floyd--Warshall algorithm.

\subsection{Principle}\label{Principle}

First, we discuss the boolean vector-matrix multiplication as follows,
\begin{equation}
	a_{i,j}^{(2)}=\sum_{k=0}^{n-1}a_{i,k}a_{k,j}=\sum_{l=0}^{n-1}\alpha[l] \wedge \beta[l],\label{a2}
\end{equation}
which propose in \cite{feng2023novel}. Formula \label{a2} describes the process of searching for a path with a length of $2$ on an unweighted graph. 

As the weight of the shortest path in each step is unknown, we consider the combination of any two paths,
\begin{equation}
	a_{i,j}=\min[(a_{i,0}+a_{0,j}),(a_{i,1}+a_{1,j}) \cdots (a_{i,n-1}+a_{n-1,j})],\label{min}
\end{equation}
which is similar to the Floyd algorithm, as it still seeks local optimal solutions. However, the following method differs from both Floyd and Dijkstra's algorithms. We will discuss the detailed comparison in Section \ref{discuss}. 

We consider fixing a certain segment of the path and then searching for possible paths through the out-degree of the endpoint node, and get,
\begin{align}
	a_{s,k,j}&=\min[(a_{s,k}+a_{k,0}),(a_{s,k}+a_{k,1}) \cdots (a_{s,k}+a_{k,n-1})],\\
	a_{s,k,j}&=\min(a_{k,0},a_{k,1},a_{k,2},a_{k,3}, \cdots a_{k,n-1}), \label{minval}
\end{align}
where $a_{s,k,j}$ represents the shortest paths from node $s$ to the any other nodes through $k$. Formula \ref{minval} implies that after determining the optimal solution at each step, we can find the shortest path by sorting the weight of the edges connected from node $k$ to the other nodes. 

However, in practice, we cannot be certain that the previous step was the optimal solution, which assumes that the path from node $s$ to $k$ is the optimal solution. In contrast, DAWN discovers only the shortest path at each step on an unweighted graph. This issue is a key limitation of DAWN for weighted graphs.

We dynamically update the shortest path from node $s$ to $s$ by traversing all nodes in the graph, and get,
\begin{equation}
	A_s=\begin{bmatrix} 
		a_{s,0,0}&,a_{s,0,1}&,\cdots&,a_{s,0,n-2}&,a_{s,0,n-1}& \\
		a_{s,1,0}&,a_{s,1,1}&,\cdots&,a_{s,1,n-2}&,a_{s,1,n-1}& \\
		\vdots&      &\ddots&          &\vdots&\\
		a_{s,n-2,0}&,a_{s,n-2,1}&,\cdots,&a_{s,n-2,n-2}&,a_{s,n-2,n-1}& \\
		a_{s,n-1,0}&,a_{s,n-1,1}&,\cdots&,a_{s,n-1,n-2}&,a_{s,n-1,n-1}& \\
	\end{bmatrix}
	\label{unweighted}
\end{equation}
and then we can get the formula as follows,
\begin{align}
	a_{s}=&[\min_{0 \leq k \leq n-1}a_{s,k,0}, \min_{0 \leq k \leq n-1}a_{s,k,1},\nonumber\\
	&\cdots \min_{0 \leq k \leq n-1}a_{s,k,n-2},\min_{0 \leq k \leq n-1}a_{s,k,n-1}].
\end{align}
Formula \ref{minval} gets the results shown in the each row of matrix $A_s$, and the shortest path from node $s$ to $j$ is the minimum of element in the each column of matrix $A_s$.

Thus, we have effectively demonstrated why the weighted version of DAWN can compute correctly, which yields the same recurrence matrix as the Floyd--Warshall algorithm. However, the time complexity is $O(n^2)$, and our work focuses on minimizing unnecessary computations as much as possible.

\subsection{GSVM}
Due to the different method of computing the recurrence matrix, the weighted version of DAWN can utilize a CSR (Compressed Sparse Row) matrix to avoid traversing unnecessary data. We present the processing flow of GSVM, 

\begin{enumerate}
	\item If path from $s$ to $j$ is exit, GSVM calculates the new path; [Line 5]
	\item If the out-degree of node $j$ is 0, GSVM does not calculate the path through node $j$; [Line 8]
	\item If path from $s$ to $k$ through $j$ is shorter than the exit path from $s$ to $k$, GSVM update the path; [Line 12]
	\item If GSVM update the path in this step, GSVM will continue, until GSVM does not update any path; [Line 22] 
\end{enumerate}
and Algorithm \ref{GSVM} describes the weighted version of the DAWN algorithm.
\begin{algorithm}[htbp]
    \caption{General Sparse Vector-Matrix Operation}
    \label{GSVM}
    \SetAlgoLined
    \LinesNumbered
    \KwIn{ A\_CSR,source, $\alpha$}
    \KwOut{$\alpha$}
    \While{step $<$ n}
    {
        step $\Leftarrow$ step + 1 \;
        ptr $\Leftarrow$ false \;
        \For{j $\in$ [0,n-1] }{
            \If{$\alpha$[j]}{
				start $\Leftarrow$ A\_CSR.row\_ptr[j] \;
                end $\Leftarrow$ A\_CSR.row\_ptr[j + 1] \;
                \If{(start $\neq$ end)}{
                    \For{k $\in$ [start,end-1]}
                    {
						index $\Leftarrow$ A\_CSR.col[k] \;
                        value $\Leftarrow$ A\_CSR.value[k] \;
						\If{($\alpha$[index] $>$ $\alpha$[j]+ value) and (index $\neq$ source)}
                        {
                            $\alpha$[index] $\Leftarrow$ $\alpha$[j]+ value\;
                            \If{(ptr = false)}
                            {ptr $\Leftarrow$ true \;}
                        }
                    }
                }
            }
        }
        \If{ptr = false}
        {break\;}
    }
    \Return{$\alpha$\;}
\end{algorithm}

In Algorithm \ref{GSVM}, $\alpha$ is the distance vector of the source to the other nodes, A\_CSR is the CSR format adjacency matrix of graph, $n$ is the number of nodes in the graph. Obviously, Algorithm \ref{GSVM} requires less than $O(n^2)$ time to solve the SSSP tasks, and we will discuss the upper bound of time complexity in next section. 

\subsection{Time Complexity of GSVM}\label{time}

Feng et al. utilized a special graph with uniformly distributed shortest paths to calculate the upper bound of DAWN's time complexity \cite{feng2023novel}, and we also employ this method to evaluate the time complexity of the weighted version. 

The shortest paths in a graph follow a uniform distribution, which means that in a specific graph, DAWN will fill the result vector with the shortest path within $Step(s)$ loops, and the number of updated shortest paths at each loop is the same. Clearly, this specific graph is a connected graph, which represents the worst-case scenario for time complexity of DAWN. The time required to run DAWN on any other graph will be less than that on this specific graph.

We can get the function of operations as follows,
\begin{align}
	f(1) & =\frac{n-pn}{Step(s)-1} \cdot d^+(j),                                                                    \\
	f(x) & = \frac{n-pn}{Step(s)-1}\cdot x \cdot \big[pn \cdot \frac{n-\frac{f(x-1)}{pn}}{n}\big], (x>1) \\
	T(n) & = \sum_{x=1}^{Step(s)-1} f(x) ,\label{FO}
\end{align}
where $n<f(x)<m$, $d^+(j)$ represents out-degree of node $j$ and $Step(s)$ represents the maximum step of the shortest path from node $s$ to any node reachable from $s$. We do not consider the case of $Step(s) <3$, which requires $O(d^+(j) \cdot [n-d^+(j)])$ time. 

We can evaluate operations using Formula \ref{FO},
\begin{align}
	T(n)= & \sum_{x=1}^{Step(s)-1}\frac{n-pn}{Step(s)-1}\cdot x \cdot \big[pn \cdot \frac{n-\frac{f(x-1)}{pn}}{n}\big], \\
	T(n)= & \frac{1-p}{Step(s)-1}\cdot \sum_{x=1}^{Step(s)-1} x \cdot \big[m -f(x-1)\big],\label{F1}
\end{align}
where $p$ represents the average connection probability of graph. We assume GSVM will always find the shorter path and update $\alpha[j]$ at the end of vector. Hence, we can simplify Formula \ref{F1} as follows,
\begin{align}
	T(n) & <\frac{1-p}{Step(s)-1}\cdot m \cdot \sum_{x=1}^{Step(s)-1} x,                 \\
	T(n) & <\frac{1-p}{2}Step(s)m.
\end{align}
Hence, DAWN requires $O(\frac{1-p}{2}\cdot Step_{max} \cdot n \cdot m)$ and $O(\frac{1-p}{2}\cdot Step(s) \cdot m)$ time for APSP and SSSP problem, respectively. 

The implementation of GSVM involves computing an increasing number of paths as the number of steps increases. Once a path is updated, it is unclear which other paths may be affected, and GSVM must recompute all paths. In the following section, we will discuss optimization methods for GSVM.

We can calculate the time complexity as follows,
\begin{align}
	g(1) & =\sum_{i=0}^{d^+(s)}d^+(i),      \\
	g(x) & =\sum_{i=0}^{nnz(\alpha)}d^+(i), \\
	T(n) & =\sum_{x=1}^{Step(s)-1}g(x),
\end{align}
where $nnz(\alpha)$ is the number of nonzero elements in $\alpha$ (Refer to Algorithm \ref{GSVM}), and $d^+(i)$ represents the out-degree of node $i$. It is not feasible to estimate the value of $nnz(\alpha)$ due to its dynamic nature and unpredictable changes.

\subsection{General Optimized Vector-Matrix Operation}

We propose the GOVM (General Optimized Vector-Matrix Operation) implementation of DAWN on weighted graphs. We can express the shortest path in each step as follows,
\begin{equation}
	p_{i,j}=\min[(p_{i,0}+a_{0,j}),(p_{i,1}+a_{1,j}) \cdots (p_{i,n-1}+a_{n-1,j})],\label{min}
\end{equation}
where $p_{i,j}$ represents the weighted of the shortest path from node $i$ to $j$, and $a_{i,j}$ represents the weighted of edge between node $i$ and $j$. 

This implies that we can optimize the algorithm by keeping track of the positions of the paths updated in the current iteration, and only considering the new paths generated from these positions in the next iteration. Algorithm \ref{GOVM} describes the optimized method of the GSVM.
\begin{algorithm}[htbp]
    \caption{General Sparse Vector-Matrix Operation}
    \label{GOVM}
    \SetAlgoLined
    \LinesNumbered
    \KwIn{ A\_CSR, source, $\alpha$, $\beta$, $\delta$}
    \KwOut{$\alpha$}
    \While{step $<$ n}
    {
        step $\Leftarrow$ step + 1 \;
        ptr $\Leftarrow$ false \;
        \For{j $\in$ [0,n-1] }{
			start $\Leftarrow$ A\_CSR.row\_ptr[j] \;
            end $\Leftarrow$ A\_CSR.row\_ptr[j + 1] \;
            \If{($\delta$[j] = true) and (start $\neq$ end)}{
                    \For{k $\in$ [start,end-1]}
                    {
						index $\Leftarrow$ A\_CSR.col[k] \;
                        value $\Leftarrow$ A\_CSR.value[k] \;
                        \If{($\alpha$[index] $>$ $\alpha$[j]+ value) and (index $\neq$ source)}
                        {
                            $\alpha$[index] $\Leftarrow$ $\alpha$[j]+ value\;
							$\beta$[index] $ \Leftarrow$ true \;
                            \If{(ptr = false)}
                            {ptr $\Leftarrow$ true \;}
                        }
                    }
            }
        }
        \If{ptr = false}
        {break\;}
		\tcp{Copy $\beta$ to $\delta$ }
		$\delta \Leftarrow \beta$ \;
    }
    \Return{$\alpha$\;}
\end{algorithm}

We record the positions of the paths updated in the current iteration in the $\beta$ vector and update the paths generated from these positions in the next iteration. As a result, we can derive the time complexity formula for GOVM as follows:
\begin{align}
	T(n) = \sum_{j=0}^{S_{wcc}(s)}d^-(j) \cdot \mu(s) = \mu(s) \cdot E_{wcc}(s),
\end{align}
where $d^-(j)$ is the in-degree of node $j$. $S_{wcc}(s)$ and $E_{wcc}(s)$ represent the size of the largest WCC (Weakly Connected Component) that node $s$ belongs to, in terms of the number of nodes and edges contained in the component, respectively. $\mu(s)$ represents the number of path update of node $s$.

The time complexity of DAWN for APSP tasks is determined by the largest WCC in the graph,
\begin{align}
	T(n) = \mu \cdot S_{wcc} \cdot E_{wcc},
\end{align}
where $S_{wcc}$ and $E_{wcc}$ denote the number of nodes and edges included in the largest WCC, $\mu$ represents the average number of path update.

In summary, DAWN based on GOVM requires $O(\mu \cdot n \cdot m)$ and $O(\mu \cdot S_{wcc} \cdot E_{wcc})$ time on the connected and unconnected graphs for APSP tasks, while require $O(\mu \cdot m)$ and $O(\mu \cdot E_{wcc})$ time for SSSP tasks, respectively.

The value of $\mu$ cannot be accurately estimated, but in Section \ref{mu}, we provided experimental validation methods that can be used to help confirm the estimated value of $\mu$ on test graphs. We selected a diverse range of graphs to ensure the generality of the empirical value obtained from these experiments. However, it should be noted that this value may not be completely applicable to all graphs. 

\subsection{Discuss}\label{discuss}
The Floyd algorithm is known for its concise code structure and mainly relies on the identification of intermediate nodes to search for possible combinations of paths to form the shortest path. The essential difference between the weighted version of DAWN and Floyd is that DAWN searches for the shortest path in terms of the steps, which is consistent with the matrix multiplication-based shortest path algorithm.

In each iteration of DAWN, the path with the least number of steps from node i to node j is first discovered. In weighted graphs, there may exist paths with more steps but lower weights, such as negative-weighted edges in a weighted graph. The weighted version of DAWN continuously searches for paths with more hops but smaller weights during the iteration, and stops when the weights of all shortest paths no longer change. This operation is similar to edge-traversal algorithms such as Dijkstra's algorithm.

The essential difference between DAWN and Dijkstra' algorithm is that DAWN does not care about the order in which paths appear. Dijkstra' algorithm requires the use of priority queues to ensure that unnecessary calculations are avoided, and its performance is often limited by the performance of the priority queue. $\Delta$-stepping implementation and other optimization methods are all focused on priority queue optimization. 

The weighted version of DAWN mainly uses simplified matrix multiplication, which visits all possible combinations of paths in each iteration and updates the optimal path combination. Each thread accesses paths from different starting points, and there is no need to update the same path simultaneously. This is also the reason why DAWN has high parallelism and scalability.


Compared to $\Delta$-stepping implementation, the weighted version of DAWN can handle negative-weighted graphs (with or without negative-weighted cycles) and has better parallelism and scalability. Compared to Floyd--Warshall algorithm, the weighted version can handle the SSSP problem and has a better time complexity in solving APSP problem.

In Section \ref{mu}, we will validate and discuss the impact of the $\mu$, which will further verify the time complexity differences between DAWN and $\Delta$-stepping implementation.


\section{Results}

In the experimental section, we mainly compared the performance difference between the SSSP API from Gunrock ($\Delta$-stepping implementation of Dijkstra's algorithm) and the weighted version of DAWN. The experiments tested the performance of the solutions on the APSP and MSSP problems.

\subsection{Experiment Introduction}
In the experiments, we used 36 general graphs and 12 large-scale graphs from the widely used open-source datasets, SNAP and SuiteSparse Matrix Collection \cite{snapnets,Davis2011matrix}. Table \ref{ExperimentalGraphs} provides the parameters of the experimental graphs, and we provide the link to the \href{https://www.scidb.cn/s/6BjM3a}{[Dataset]}. Table \ref{TestMachine} shows the parameters of the test machine. Table \ref{Algorithms} shows the solutions of the algorithms.

\begin{table}[h]
	\begin{center}
		\caption{Parameters of the Experimental Graphs (Nodes and Edges)}\label{ExperimentalGraphs}%
		\begin{minipage}{\linewidth}
			\begin{tabular*}{\textwidth}{@{\extracolsep{\fill}}ccccc@{\extracolsep{\fill}}}
				\toprule
				\multicolumn{5}{c}{\textbf{\large Nodes}}\\
				\midrule
				< 100K & 100K $\thicksim$ 200K & 200K $\thicksim$ 1M & 1M $\thicksim$ 100M& > 100M \\
				\midrule
				14 & 22 & 9&11& 3 \\
				\bottomrule
			\end{tabular*}
		\end{minipage}
		\begin{minipage}{\linewidth}
			\begin{tabular*}{\textwidth}{@{\extracolsep{\fill}}ccccc@{\extracolsep{\fill}}}
				\toprule
				\multicolumn{5}{c}{\textbf{\large Edges}}\\
				\midrule
				< 1M & 1M $\thicksim$ 5M & 5M $\thicksim$ 20M & 20M $\thicksim$ 100M & > 100M \\
				\midrule
				8 & 16 & 23 & 6 & 6 \\
				\bottomrule
			\end{tabular*}
		\end{minipage}
	\end{center}
\end{table}

\begin{table}[h]
	\begin{center}
		\begin{minipage}{\linewidth}
			\caption{Parameters of the Test Machine}\label{TestMachine}%
			\begin{tabular*}{\textwidth}{@{\extracolsep{\fill}}ccc@{\extracolsep{\fill}}}
				\toprule
				Hardware & Machine1&Machine2\\
				\midrule
				CPU	& Intel Core i5-13600KF& AMD EPYC Milan 7T83\\
				RAM	& 32GB & 128GB\\
				GPU	& NVIDIA RTX 3080TI & --\\
				Compiler& NVCC and GCC 9.4.0 & GCC 9.4.0\\
				OS	 &  Ubuntu 20.04 & Ubuntu 20.04\\
				Toolkit	& CUDA 12.0& --\\
				\bottomrule
			\end{tabular*}
		\end{minipage}
	\end{center}
\end{table}

\begin{table}[htbp]
	\begin{center}
		\begin{minipage}{\linewidth}
			\caption{Parameters of the Algorithms}\label{Algorithms}%
			\begin{tabular*}{\textwidth}{@{\extracolsep{\fill}}cl@{\extracolsep{\fill}}}
				\toprule
				Abbreviation &	Solution\\
				\midrule
				GDS             & The  $\Delta$-stepping implementation on GPUs,\\
				&  which is from the Gunrock \cite{Wang:2017:GGG,Osama:2022:EOP,10.1145/3572848.3577434}\\
				DAWN	        & DAWN with RTX 3080TI\\
				DAWN(20)	        & DAWN with I5-13600KF \\
				DAWN(64)	        & DAWN with EPYC 7T83 using 64 threads \\
				\bottomrule
			\end{tabular*}
		\end{minipage}
	\end{center}
\end{table}

Gunrock is a CUDA library that has been specifically developed for graph-processing on the GPU. This library has been designed with the aim of achieving a superior balance between performance and expressiveness, which has been accomplished through the integration of high-performance GPU computing primitives and optimization strategies, with a particular emphasis on fine-grained load balancing \cite{Wang:2017:GGG,Osama:2022:EOP,10.1145/3572848.3577434}.

The weighted version of DAWN is a component of Gunrock, and the implementation code can be obtained from the main branch of the \href{https://github.com/gunrock/gunrock}{[Gunrock]}. Furthermore, we have provided the solution that does not rely on the Gunrock, which enables researchers who lack access to the NVCC compiler to utilize DAWN. \href{https://github.com/lxrzlyr/DAWN-An-Noval-SSSP-APSP-Algorithm}{[GitHub]}

\subsection{Validation Method}\label{mu}

We conducted two comparative experiments to determine the range of the $\mu$:
\begin{enumerate}
	\item The baseline experiment sets the weights of the graph to 1.0f.
	\item The comparative experiment sets the weights of the graph in range [0.0f, 2.0f], randomly.
\end{enumerate}

\begin{figure}[htbp]
	\begin{minipage}{\linewidth}
		\centering
		\includegraphics[width=\linewidth]{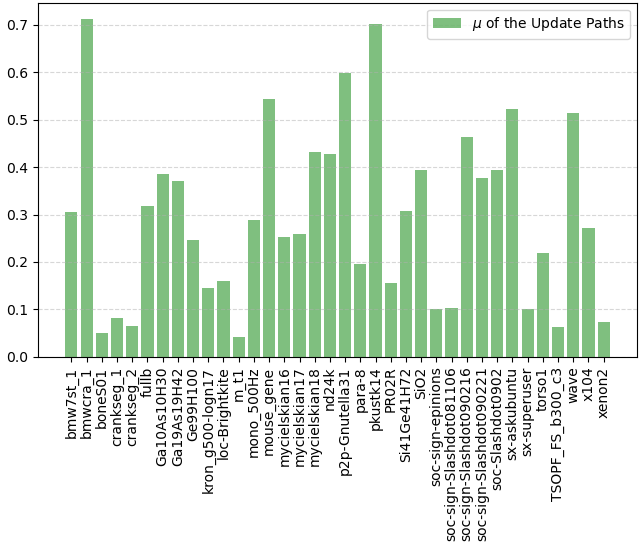}
		\caption{Evaluation of Weighted Version on 36 general graphs: Ratio of Updated Paths (Present the portion of $\mu$ that has increased)}
		\label{fig:mu}
	\end{minipage}
\end{figure}

In the baseline experiment group, once the shortest path is discovered, it will not be updated. The time difference between the comparative experiment group and the baseline experiment group is caused by the increase in computation resulting from path updates. Figure \ref{fig:mu} illustrates the proportion of updated paths in the weighted version of DAWN on a weighted graph. 

On 36 general graphs, the weighted version of DAWN algorithm updated an average of 29.578\% of the paths, with range of [4.279\%, 71.118\%]. These results indicate that the mean value of $\mu$ is equal to 1.296, which is dependent on the structure of the graphs, such that even changing the edge weights on the same graph can affect its value.

Experimental results have demonstrated that $\mu$ is a constant value independent of the number of nodes and edges in the graph. Furthermore, both the weighted version of DAWN and GDS algorithms exhibit similar time complexity on the general graphs. These findings provide a solid foundation for subsequent performance evaluations.

\subsection{Scalability}\label{scalability}

\subsection{Speedup}

\subsection{Latency}

\subsection{Large-scale Graph}\label{large}

\section{Conclusion}
In this paper, we propose a weighted version of DAWN, which requires $O(\mu \cdot n \cdot m)$ and $O(\mu \cdot S_{wcc} \cdot E_{wcc})$ time on the connected and unconnected graphs for APSP problems, respectively. The weighted version of DAWN requires $O(\mu \cdot m)$ and $O(\mu \cdot E_{wcc})$ time on the connected and unconnected graphs for SSSP problems.

The experimental results demonstrate that DAWN exhibits significant performance improvement on high-performance processors and surpasses Gunrock's performance even on low-cost processors. DAWN demonstrates superior performance and broader applicability over GDS on large-scale graphs.

The high scalability and high parallelism of DAWN provide a solution for the analysis of the shortest paths problem in various graphs. We will develop the version of DAWN running on the weighted graph in the future. We hope to take the DAWN from a promising proof-of-concept to a practical tool that can be used for real-world graph analysis applications in the future.


\begin{acks}
	To Robert, for the bagels and explaining CMYK and color spaces.
\end{acks}

\bibliographystyle{ACM-Reference-Format}
\bibliography{bibfile}


\end{document}